\documentclass{ifacconf}

\usepackage{graphicx}      % include this line if your document contains figures
\usepackage{natbib}        % required for bibliography

\usepackage{graphicx}
\usepackage{amsmath}
\usepackage{amssymb}
\usepackage{calc}
\usepackage{multirow}
\usepackage{mathrsfs}
\usepackage{enumitem}
\usepackage{float}
\usepackage{cool}
\usepackage{url}
\usepackage{nicefrac, xfrac}
\usepackage{tikz}
\usepackage{times}
\usepackage{color,colortbl}
\usepackage{tikz}
\usetikzlibrary{calc,arrows,positioning}
\usepackage{enumitem}
\usepackage{commath}
\usepackage{tensor}
\usepackage{mathtools}
\usepackage{amsmath,amsfonts,amssymb,commath}
\usepackage{gensymb}
\usepackage{siunitx}
\usepackage{pifont}
\newcommand{\bff}{\ensuremath{\mathbf{f}}}

\newcommand{\bfh}{\ensuremath{\mathbf{h}}}

\newcommand{\bfv}{\ensuremath{\mathbf{v}}}
\newcommand{\bfw}{\ensuremath{\mathbf{w}}}
\newcommand{\bfx}{\ensuremath{\mathbf{x}}}

\newcommand{\bfz}{\ensuremath{\mathbf{z}}}

\newcommand{\bfC}{\ensuremath{\mathbf{C}}}

\newcommand{\bfJ}{\ensuremath{\mathbf{J}}}

\newcommand{\bfQ}{\ensuremath{\mathbf{Q}}}

\newcommand{\bbfx}{\ensuremath{\bar{\mathbf{x}}}}

\newcommand{\hbfx}{\ensuremath{\hat{\mathbf{x}}}}

\DeclareMathOperator*{\mean}{E}				% Expectation operator
\DeclareMathOperator*{\cov}{cov}			% Covariance operator
\DeclareMathOperator*{\var}{var}			% Variance operator
\newcommand{\bfDelta}{\ensuremath{\boldsymbol{\Delta}}}

\newcommand{\bfxi}{\ensuremath{\boldsymbol{\xi}}}

\RequirePackage{xspace}

\def\calO{\mathcal{O}}
\def\bfXi{\mathbf \Xi}

\def\real{\mathbb{R}}

\def\bfxi{\ensuremath{\boldsymbol{\xi}}}
\begin{document}

\thispagestyle{empty}

\begin{center}
\vspace*{3cm}
{\Large\bfseries Preprint Notice}
\vspace{1cm}

This manuscript is a preprint.\\[0.5em]
The final version is currently under review for IFAC WC 2026 with IFAC Journal of Systems and Control option. Please cite the published version if accepted.

\end{center}

\clearpage

\setcounter{page}{1}

\begin{frontmatter}

\title{Lagrangian Grid-based Estimation of Nonlinear Systems with Invertible Dynamics}%\thanksref{footnoteinfo}} 

\thanks[footnoteinfo]{J. Dun\'ik, J. Matou\v{s}ek, J. Krejčí, and M. Brandner were partially supported by the Czech Science Foundation (GACR) under grant GA 25-16919J.}

\author[First]{Jind\v{r}ich Dun\'ik, Jan Krej\v{c}\'i} 
\author[Second]{, Jakub Matou\v{s}ek} 
 \author[Third]{, Marek Brandner} 
 \author[Fourth]{Yeongkwon Choe}
 \author[Fifth]{, Chan Gook Park}

 \address[First]{Dept. of Cybernetics, Univ. of West Bohemia in Pilsen, Czech Republic (e-mail: \{dunikj, jkrejci\}@kky.zcu.cz)}
 \address[Second]{Oden Institute, The University of Texas at Austin, TX, USA (e-mail: jakub.matousek@austin.utexas.edu)}
 \address[Third]{Dept. of Mathematics, Univ. of West Bohemia in Pilsen, Czech Republic (e-mail: brandner@kma.zcu.cz).}
 \address[Fourth]{Dept. of Mechatronics Eng., Kangwon National Univ.
       Chuncheon, Republic of Korea (e-mail: ychoe@kangwon.ac.kr)}
\address[Fifth]{Dept. of Aerospace Eng./ASRI, Seoul National Univ.,
       Seoul, Republic of Korea (e-mail: chanpark@snu.ac.kr)}

\begin{abstract}                % Abstract of 50--100 words
This paper deals with the state estimation of non-linear and non-Gaussian systems with an emphasis on the numerical solution to the Bayesian recursive relations.
% In particular, the stress is laid on the design of the grid-based filter (GbF) following the recently developed Lagrangian perspective for linear systems also for the systems with nonlinear, but invertible, dynamics.
In particular, this paper builds upon the Lagrangian grid-based filter (GbF) recently-developed for linear systems and extends it for systems with nonlinear dynamics that are invertible.
The proposed nonlinear Lagrangian GbF reduces the computational complexity of the standard GbFs from quadratic to log-linear, while preserving all the strengths of the original GbF such as robustness, accuracy, and deterministic behaviour. The proposed filter is compared with the particle filter in several numerical studies using the publicly available MATLAB\textregistered\ implementation.%\footnote{\url{https://github.com/pesslovany/Matlab-LagrangianPMF}}.
%The proposed LGbF has been implemented in \textsc{MATLAB}{\textregistered}, and the code has been made publicly available\footnote{\textcolor{red}{TBD UPD}%\url{https://github.com/pesslovany/Matlab-LagrangianPMF-simulated-smoothing}}.
\end{abstract}


\end{frontmatter}
%===============================================================================

\section{Introduction}

State estimation is essential in numerous areas of modern society, including navigation, timekeeping, speech and image processing, fault detection, optimal control, and tracking. Consequently, it has been a subject of considerable research interest for several decades.  

This paper deals with Bayesian state estimation for discrete-time stochastic state-space models. The emphasis is laid on global filters based on the numerical solution to the \emph{Bayesian recursive relations} (BRRs) that allow calculation of the probability density function (PDF) of the system state conditioned on the measurements. Although the global\footnote{In contrast, local filters such as the extended or unscented Kalman filters \citep{BaLiKi:01}, are computationally efficient but they provide reasonable estimation performance for mildly nonlinear or non-Gaussian models.} filters are computationally demanding, they provide reliable and consistent estimates even for highly nonlinear and/or non-Gaussian models, which can be found in areas such as the terrain-aided navigation, multitarget tracking, economics, or weather forecasting.
%in which the FRRs lack closed-form solutions and local filters—such as the extended or unscented Kalman filter—fail to achieve acceptable performance \citep{MaDuSt:25}.  

Two main numerical strategies for solving the BRRs have been developed in the literature: approaches based on \emph{stochastic} and on \emph{deterministic} integration rules. The former relies on Monte Carlo (MC) integration and leads to the family of particle filters (PFs) \citep{DoFrGo:01}. The latter employs deterministic numerical integration rules, for example the midpoint rule, and results in the grid-based filters (GbFs) \citep{SiKraSo:06}. {In this paper, we focus on the latter \textit{grid-based} approach that offers (\textit{i}) an increased resilience to initialisation errors and measurement outliers due to the grid design allowing to cover a significant part of the state space, and (\textit{ii}) deterministic behaviour\footnote{GbFs produce for identical data the identical estimates across repeated runs and computing platforms.} being required by safety-critical applications \citep{AnHa:10, MaDiLiFa:23, AnHa:06,OrSkToGu:10}}. In the GBF design, two principal approaches can be found; standard approach following the \textit{Eulerian} integration philosophy developed in sixties \citep{BuSe:71,SiKraSo:06}, and modern approach following the \textit{Lagrangian} philosophy developed in this context in early 2010s \citep{PaZh:11,MaDuSt:25}.

%\textcolor{blue}{Marek: correct the sentence please :-) Jindra: třeba nějak takhle? Nevím ale, jestli je to srozumitelné. Není to ale samozřejmě úplně přesné, protože přechod mezi EGbF a LGbF je svým způsobem "trochu spojitý".} 
Standard Eulerian GbFs, known as the point-mass filters, are based on such grid design for the numerical BRRs solution, that does \textit{not} closely follow the state dynamics (they are based on a fixed grid or an adaptive grid, where adaptivity is not directly linked to the state dynamics; nor do they use any other form of adaptivity controlled by state dynamics). Although these filters are conceptually simpler and can be designed for an arbitrary nonlinear and/or non-Gaussian model, they suffer from the computational complexity $\mathcal{O}(N^{2})$ \citep{Be:99,SiKraSo:06}, where $N$ is the the number of grid points used for the numerical solution. On the other hand, the Lagrangian GbFs (LGbFs) take advantage of the grid dynamics being identical with the state dynamics. To allow this behaviour of the grid, the prediction step of the LGbF is split into two subsequent operations; advection (respecting the state function) and diffusion (respecting the state noise).
% As a consequence, the LGbF are much more computationally efficient (despite more complex grid manipulation) with the complexity of $\mathcal{O}(N\log_{2}N)$, but they have been designed for models with linear\footnote{Linear models are quite popular in tracking, where the object dynamics can be described by nearly constant velocity/acceleration model, constant turn rate model, or the Singer model \citep{BaLiKi:01}.} dynamics only.
{As a consequence, the LGbFs are much more computationally efficient (despite more complex grid manipulation) with the complexity of $\mathcal{O}(N\log_{2}N)$. The LGbFs have been designed mostly for a models with linear\footnote{Linear models are quite popular in tracking, where the object dynamics can be described by nearly constant velocity/acceleration model, constant turn rate model, or the Singer model \citep{BaLiKi:01}.} dynamics \citep{MaDuSt:25}. However, as the assumption of linear dynamics can be limiting for certain applications, the concept of the LGbF has been extended for the models with nonlinear not only the measurement equation, but also the state equation \citep{DuMaSt:23}. This concept is based on evaluation of the conditional PDFs on \textit{two} grids; the coarse grid used in the filter time recursion and the fine grid redefined at each time instant allowing more reliable and accurate transformation of the filtering PDF through the nonlinear state function. However, for dynamics with a higher degree of nonlinearity \citep{DuMaSt:23}, this approach can lead to either (\textit{i}) a fragmented conditional PDF, where certain grid points are incorrectly assigned with a zero probability, or (\textit{ii}) definition of the sufficiently dense fine grid, which leads to significantly increased and hardly predictable complexity (from the perspective of computational power and also memory requirements).
}

{The goal of this paper is to propose the single-grid LGbF that does not use a fine grid potentially leading to the fragmented conditional PDF or increased computational resources. This LGbF is proposed for nonlinear state-space models with invertible dynamics, which are typical in the areas of tracking, economy, and physics to name a few \citep{He:76,BaLiKi:01}. The basic idea of the proposed LGbF stems from the design of an grid with \textit{N} non-equidistantly placed grid points in the prediction step \textit{strictly} respecting the inverse state dynamics.} Such a design of the single grid then allows \textit{efficient} solution of the prediction step by the advection and diffusion operations, which remains identical with the LGbF for the linear case. It means that the introduced solution can easily be plugged into any implementation of the LGbF. The proposed LGbF is illustrated in set of numerical studies and the codes are on GitHub.

%In particular, two models are treated; models with analytically invertible and non-invertible dynamics. Whereas the former allows 
 
The rest of the paper is organized as follows. Section 2 introduces Bayesian estimation of nonlinear state-space model with an introduction to GbF design. In Section 3, the LGbF is proposed and the algorithm is illustrated. Numerical evaluation is given in Section 4 and conclusion is drawn in Section 5.

\section{Model and Bayesian Estimation}
The following discrete-time state-space model of a nonlinear stochastic dynamic system with additive noises
\begin{align}
\bfx_{k+1}&=\bff_{k}(\bfx_{k})+\bfw_{k}, \label{eq:asx}\\
\bfz_{k}&=\bfh_k(\bfx_k)+\bfv_k,\label{eq:asz}
\end{align}
is considered, where the vectors $\bfx_k\in\real^{n_x}$, and $\bfz_k\in\real^{n_z}$ represent the \textit{unknown} state of the model and \textit{known} measurement at time instant $k$, respectively. The state function $\bff_k:\real^{n_x}\rightarrow\real^{n_x}$, and measurement function $\bfh_k:\real^{n_x}\rightarrow\real^{n_z}$ are supposed to be \textit{known}. Particular realizations of the state and measurement noises $\bfw_k$ and $\bfv_k$ are \textit{unknown}, but their PDFs, i.e., the state noise PDF $p_{\bfw_k}(\bfw_k)$ and the measurement noise PDF $p_{\bfv_k}(\bfv_k)$, are supposed to be \textit{known}, as well as the initial state PDF $p_{\bfx_0}(\bfx_0)$. The noises and initial state are independent. Note that whenever it does not compromise clarity, the subscripts are omitted from the PDF notation, i.e., $p(\bfw_k)=p_{\bfw_k}(\bfw_k)$.

\vspace*{0mm}
\hspace*{-0mm}\textbf{Assumption:} The state dynamics, defined by the function $\bff_{k}$ is analytically invertible, i.e., the inverse $\bff_k^{-1}:\mathbb{S}\rightarrow \mathbb{S}$ of $\bff_k$ with the property $\bff_k^{-1}(\bff_k(\bfx_k))=\bfx_k, \forall \bfx_k\in\mathbb{S}$, exists, where $\mathbb{S}$ stands for the considered region of the state space.

\vspace*{0mm}
\hspace*{-0mm}\textbf{Note:} State-space model \eqref{eq:asx}, \eqref{eq:asz} with invertible function $\bff_k$ can be found in the areas of navigation, tracking, economy, physics, and biology. These models include, for example, the coordinated turn model with unknown turn rate or {Dubins model} used in targed tracking, or Chirikov–Taylor map or Hénon map for description of chaotic systems.

\subsection{Bayesian Estimation}
% The main goal of state estimation in the Bayesian framework is to find an estimate of the PDF of the state $\bfx_k$ conditioned on all measurements $\bfz^l=[\bfz_0,\bfz_1,\ldots,\bfz_l]$  up to the time instant $l$, i.e., the conditional PDF $p(\bfx_k|\bfz^l), \forall k$, is sought.
The main goal of Bayesian state estimation is to provide an accurate approximation of the PDF of the state $\bfx_k$ conditioned on all measurements $\bfz^l=[\bfz_0,\bfz_1,\ldots,\bfz_l]$  up to the time instant $l$, i.e., the conditional PDF $p(\bfx_k|\bfz^l), \forall k$, is sought.
If $k=l$, then the estimation task is \textit{filtering}, if $k>l$, then we talk about  \textit{prediction}.

\subsection{Bayesian Recursive Relations}
%The general solution to the state estimation is given by the BRRs for the filtering and predictive conditional PDFs computation \citep{AnMo:79}
The BRRs enables calculation of the filtering and predictive conditional PDFs \citep{AnMo:79}
\begin{align}
p(\bfx_k|\bfz^k)&=\frac{p(\bfx_k|\bfz^{k-1})p(\bfz_k|\bfx_k)}{p(\bfz_k|\bfz^{k-1})},\label{eq:filt}\\
p(\bfx_{k+1}|\bfz^{k})&=\int p(\bfx_{k+1}|\bfx_{k})p(\bfx_{k}|\bfz^{k})d\bfx_{k},\label{eq:pred}
\end{align}
where $p(\bfx_{k}|\bfz^{k-1})$ is the one-step predictive PDF computed by the Chapman-Kolmogorov equation (CKE) \eqref{eq:pred} and $p(\bfx_k|\bfz^k)$ is the filtering PDF computed by the Bayes' rule \eqref{eq:filt}. The PDFs $p(\bfx_{k+1}|\bfx_{k})=p_{\bfw_k}(\bfx_{k+1}-\bff_k(\bfx_{k}))$ and $p(\bfz_{k}|\bfx_{k})=p_{\bfv_k}(\bfz_{k}-\bfh_k(\bfx_{k}))$ are the state transition PDF obtained from \eqref{eq:asx} and the measurement PDF obtained from \eqref{eq:asz}, respectively. The PDF $p(\bfz_k|\bfz^{k-1})=\int p(\bfx_k|\bfz^{k-1})p(\bfz_k|\bfx_k)$ $d\bfx_k$ is the one-step predictive PDF of the measurement. The recursion \eqref{eq:filt}, \eqref{eq:pred} starts from $p(\bfx_0|\bfz^{-1})=p(\bfx_0)$. %For this part of the estimation routine an efficient and robust Lagrangian grid-based filter (LGbF) was recently proposed.

\subsection{Point-Mass Density: Tool Enabling Numerical BRRs Solution}
\begin{figure}[]
	\centering
	\includegraphics[width=1\linewidth]{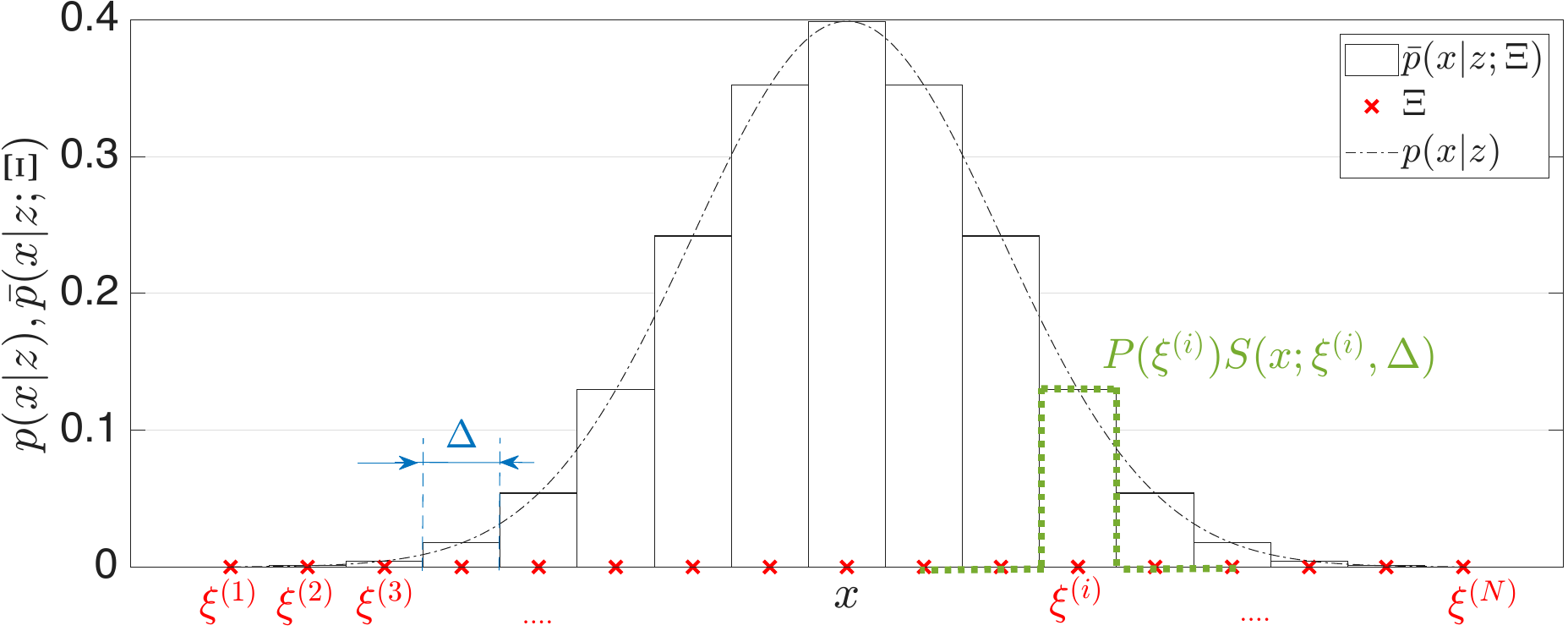}\vspace*{-2mm}
	\caption{Probability density function and its approximation by point-mass density.}
	\label{fig:PMD}\vspace*{-0mm}
\end{figure}

GbF design relies on a grid of points in the state space, on which a piecewise constant approximation of the conditional PDF, called the point-mass density (PMD), is defined and calculated iteratively. An original PDF and its PMD approximation is illustrated in Fig. \ref{fig:PMD}. The PMD approximating the conditional PDF ${p}(\bfx_k|\bfz^l)$ can be described as \citep{Be:99}
\begin{align}
    \bar{p}(\bfx_k|\bfz^l;\bfXi_k)\triangleq\sum_{i=1}^N {P}_{k|l}(\bfxi_k^{(i)}) S(\bfx_k;\bfxi_k^{(i)},\bfDelta_k^{(i)}),\label{eq:pmd}
\end{align}
where $\bfxi_k^{(i)}\in\real^{n_{\bfx}}$ is the $i$-th point of the grid $\bfXi_k=\{\bfxi_k^{(i)}\}_{i=1}^N$ with the cardinality $N = \prod_{i=1}^{n_x}N_i$, $N_i$ is the number of grid points per dimension, the weight ${P}_{k|l}(\bfxi_k^{(i)}) \propto p(\bfxi_k^{(i)}|\bfz^l)$ is the value of the PDF $p(\bfx_k|\bfz^l)$ evaluated at  $\bfxi^{(i)}_k$ and normalized w.r.t. $\forall i$, and $S(\bfx_k;\bfxi_k^{(i)},\bfDelta_k^{(i)})$ is an indicator function defining the non-overlapping neighbourhood described by the vector $\bfDelta_k^{(i)}$;  $S(\bfx_k;\bfxi_k^{(i)},\bfDelta_k^{(i)})=1$ if $\bfx_k$ is in the part of the state space defined by $\bfxi_k^{(i)},\bfDelta_k^{(i)}$, and $S(\bfx_k;\bfxi_k^{(i)},\bfDelta_k^{(i)})=0$ otherwise. 

{The grid is designed to well-cover the PDF support. There are two possibilities to design the grid; \textit{moment-based} and \textit{grid-flow}-based \citep{Be:99,SiKraSo:06}. In the \textit{former} grid design, the grid is centred at $\hbfx_{k|l}=\mean[\bfx_k|\bfz^l]$ and spans the state-space region determined by the covariance matrix $\bfC_{k|l}=\cov[\bfx_k|\bfz^l]$ and the scaling factor $\kappa$. In particular, for $i$-th element of the state vector, the grid points are designed to equidistantly cover the region $\langle\hbfx_{k|l}(i)-\kappa\sqrt{\bfC_{k|l}(i,i)},\hbfx_{k|l}(i)+\kappa\sqrt{\bfC_{k|l}(i,i)}\rangle$, where a reasonable choice of $\kappa$ is between five and six and it allows to well support even multi-modal and heavy-tailed PDFs. The grid can be also rotated using the covariance matrix $\bfC_{k|l}$ eigenvectors. In the \textit{latter} grid design, the rectangular region to be covered by the grid $\bfXi_k$ is determined by the boundaries of the grid $\bfXi_{k-1}$ ``flowing'' via the dynamics $\bff_k(\cdot)$. The algorithm proposed in this paper combines both design approaches and the used notation $\hbfx(i)$ means $i$-th element of the vector $\hbfx$ and $\bfC(i,j)$ is the element of the matrix $\bfC$ at $i$-th row and  $j$-th column.}

Following this design, the neighbourhood of $i$-th point is often \textit{hyper-rectangular} with the edges $\bfDelta_k^{(i)}=\left[\bfDelta_k^{(i)}(1), \bfDelta_k^{(i)}(2),\right.$ $\left.\bfDelta_k^{(i)}(3), \ldots,\bfDelta_k^{(i)}(n_x)\right]^T, \forall i$. Then, the volume of the neighbourhood is $\delta_k^{(i)}=\prod_{j=1}^{n_x}\bfDelta_k^{(i)}(j)$. %The notation $\bfDelta_k^{(i)}(j)$ means $j$-th element of the vector $\bfDelta_k^{(i)}$. 
If the grid points are equidistantly placed, then $\bfDelta_k^{(i)}=\bfDelta_k^{(j)}=\bfDelta_k$ and $\delta_k^{(i)}=\delta_k^{(j)}=\delta_k,\forall i,j$ %, i.e., the neighbourhood dimensions are the same for all grid points, in which case 
and the upper index is omitted for the sake of notational simplicity.%\footnote{Point-independent neighbourhood is considered for the sake of clarity.}

%The standard GbFs are implemented using column vectors of weights (vectorized representation), further denoted with double column as $P^{(:)} = [P^{(1)},...,P^{(N)} ]^T$, where one element is $P^{(i)}$

%However, the efficient Lagrangian-based implementation exploits the underlying physical nature of the state. Therefore, the weights are stored as a tensor $P \in \mathbb{R}^{N_1 \times N_2 \times ... \times N_{n_x}}$ and their element indexed as $P^{(i_1,...,i_{n_x})}$. Where the order does not matter (i.e. it does not matter wheter $P$ is a tensor or vector) the tensor weights can be also indexed by linear indexing $P^{(i)}$. For example, normalisation of the PMD can be written as $P_{k|l}^{(i)} = \frac{P_{k|l}^{(i)}}{\delta_0\sum_{i=1}^N P_{k|l}^{(i)}} $.

\subsection{Eulerian Grid-based Filter: Algorithm Summary}\label{sec:EGbF}
Design of any filter, including the standard Eulerian grid-based filter (EGbF), starts with definition of the initial PDF $p(\bfx_0|\bfz^{-1})=p(\bfx_0)$.
\subsubsection{Initialisation:}
The GbF at $k=0$ is initialized by designing the initial grid $\bfXi_0$ and the initial, i.e., the predictive, PMD $ \bar{p}(\bfx_0|\bfz^{-1};\bfXi_0)$ according to \eqref{eq:pmd}.
\subsubsection{Filtering:} The measurement update step lies in accommodation of the measurement $\bfz_k$ using the numerical solution to the Bayes' rule \eqref{eq:filt}. The $i$-th weight of the filtering PMD  $\bar{p}(\bfx_k|\bfz^{k};\bfXi_k)$ is calculated as
\begin{align}
    P_{k|k}(\bfxi_k^{(i)})\propto p(\textbf{z}_k|\bfx_k=\bfxi_k^{(i)})P_{k|k-1}(\bfxi_k^{(i)}),\label{eq:pmf_filtrace_a}
\end{align}
where $p(\textbf{z}_k|\bfx_k=\bfxi_k^{(i)})=p_{\bfv_k}(\textbf{z}_k-\bfh_k(\bfxi_k^{(i)}))$.
\subsubsection{Prediction:} The time update step starts with specification\footnote{The new grid $\bfXi_{k+1}$ location can be calculation of the predictive mean and covariance matrix by a local filter \citep{ChGo:21}. The further details on the grid design are given later.} of a ``new'' grid $\bfXi_{k+1}$ for the time step $k+1$  and a numerical solution to the CKE \eqref{eq:pred}. The $j$-th weight of the predictive PMD  $\bar{p}(\bfx_{k+1}|\bfz^{k};\bfXi_{k+1})$ is calculated as
\begin{align}
P_{k+1|k}(\bfxi_{k+1}^{(j)})=\sum_{i=1}^Np(\bfxi_{k+1}^{(j)}|\bfxi_k^{(i)})P_{k|k}(\bfxi_k^{(i)}) \delta_k, \label{eq:pmf_pred_predP}
\end{align}
where $p(\bfxi_{k+1}^{(j)}|\bfxi_k^{(i)})=p_{\bfw_k}(\bfxi_{k+1}^{(j)}-\bff_k(\bfxi_k^{(i)}))$.
Due to the evaluation of the state noise PDF $N^2$-times for all combinations of the grid points $\bfxi_{k+1}^{(j)}, \bfxi_k^{(i)}, \forall i,j,$ the prediction step is the most demanding part of the EGbF with the computational complexity $\calO(N^2)$. This is, in fact, the main disadvantage of the standard EGbFs compared to the PFs and the main motivation for the LGbF development \citep{PaZh:11,MaDuSt:25}.

\section{Lagrangian Grid-based Filter: Time Update}% Linked Closer with State Dynamics}
The LGbF reduces computational complexity of the EGbF to the same level as the PF, while retaining key advantages such as robustness and deterministic behaviour. 

The LGbFs share the initialization and measurement update with the EGbFs. The difference lies in the time update, where the ultimate goal of the Lagrangian formulation is to reduce the computational complexity of the Eulerian formulation \eqref{eq:pmf_pred_predP}. The basic idea of the Lagrangian formulation is in close interconnection of the grid dynamics, i.e., design of the grid $\bfXi_{k+1}$ based on $\bfXi_{k}$, and the state dynamics \eqref{eq:asx}, while maintaining the equidistant grid structure enabling fast solution to the convolution using the discrete Fourier transform (DFT). However, this \textit{single-grid} time update step in the Lagrangian formulation has been proposed for a \textit{linear dynamics} only \citep{MaDuSt:25}.

\subsection{Goal of the Paper}
In this paper, we extend the applicability of the efficient \textit{single-grid} Lagrangian prediction also for the nonlinear state-space models with invertible dynamics~\eqref{eq:asx}. The proposed approach naturally encompasses observable linear models (such as tracking models) used in \citep{MaDuSt:25} as a special case.

\subsection{Advection and Diffusion}
The Lagrangian approach splits the state dynamics \eqref{eq:asx}
\begin{align}
    \bfx_{k+1}=\bfx_{k+1}^{\mathrm{adv}}+\bfw_k.\label{eq:adv_dif}
\end{align}
and the GbF prediction\footnote{The EGbF prediction takes into account the function $\bff_k(\cdot)$ and the noise properties $p(\bfw_k)$ within a single operation \eqref{eq:pmf_pred_predP}.} \eqref{eq:pmf_pred_predP} into:
\begin{itemize}
    \item \emph{Advection}: respecting the ``deterministic'' component
    $\bfx_{k+1}^{\mathrm{adv}}$ $=\bff_k(\bfx_k)$ of \eqref{eq:asx}, by $\bff_k$-driven transform of the filtering PMD grid to determine the new grid $\bfXi_{k+1}$ and the advected PMD, %ing a grid used for evaluation of the filtering PMD forward in time via ,
    \item \emph{Diffusion}: respecting the ``stochastic'' component $\bfw_k$ of the state equation \eqref{eq:asx}, by convoluting the advected PMD and the $p(\bfw_k)$ on %using the convolution on the 
    $\bfXi_{k+1}$. % of the predictive PMD.% with equidistant grid points layout.
\end{itemize}

% Respecting the splitting of the prediction step to the advection and the diffusion, the state dynamics \eqref{eq:asx} can be  written in a convenient form as
% \begin{align}
%     \bfx_{k+1}=\bfx_{k+1}^{\mathrm{adv}}+\bfw_k,\label{eq:adv_dif}
% \end{align}
% where $\bfx_{k+1}^{\mathrm{adv}}=\bff_k(\bfx_k)$ stands for the deterministic component of the dynamics.

\subsection{Advection Solution}
{The key ingredient of any GbF is the design of the grid $\bfXi_{k+1}$, which supports the conditional PMDs well $\forall k$, i.e., the grid is non-degenerative. The grid shall also maintain other properties dictated by the selected prediction step design. In the LGbFs, the new grid $\bfXi_{k+1}$ has to further (\textit{i}) respect the dynamics driven by $\bff_k(\cdot)$ and (\textit{ii}) be equidistant. }

{To ensure the desired properties of the grid $\bfXi_{k+1}$, the developed approach consists of three principal operations; (\textit{i}) moment-based design of the equidistant grid $\bfXi_{k+1}$, (\textit{ii}) its propagation backward in time and interpolation of the filtering PMD on this grid, and (\textit{iii}) prediction solution using the Lagrangian approach and the DFT.} Detailed description of the needed steps for the prediction PMD calculation follows.%Design of the desired grid $\bfXi_{k+1}$ and the whole advection solution follows.

\subsubsection{Design of New Grid for Predictive PMD:}
Having the filtering PMD $\bar{p}(\bfx_k|\bfz^{k};\bfXi_k)$ \eqref{eq:pmf_filtrace_a}, the filtering mean $\hbfx_{k|k}=\mean[\bfx_k|\bfz^{k}]$ and covariance matrix $\bfC_{k|k}=\cov[\bfx_k|\bfz^{k}]$ can be calculated as
\begin{align}
    %\mean[\bfx_k|\bfz^{k}]&\approx
    \hbfx_{k|k}&\approx\sum\limits_{i=1}^{N}\delta_kP_{k|k}(\bfxi^{(i)}_{k})\bfxi^{(i)}_{k},\label{eq:mean}\\
    %\cov[\bfx_k|\bfz^{k}]&\approx
    \bfC_{k|k}&\approx\sum_{i=1}^{N}\delta_kP_{k|k}(\bfxi^{(i)}_{k})%\nonumber\\
    \left(\bfxi^{(i)}_{k}-\hbfx_{k|k}\right)\left(\cdot\right)^T\!.\label{eq:cov}
\end{align}
Then, we can calculate \textit{approximate} predictive moments $\hbfx_{k+1|k}$, $\bfC_{k+1|k}$ simply by the prediction step of any local filter, such as the extended or unscented Kalman filter \citep{BaLiKi:01}. The predictive moments specify the part of the state space, where the predictive PMD $\bar{p}(\bfx_{k+1}|\bfz^{k};\bfXi_{k+1})$ is expected to lie and where the new grid $\bfXi_{k+1}$ is designed in rectangular and equidistant layout {respecting the selected scaling factor $\kappa$}. % in that area in the usual manner. % to well cover the predictive PMD support. 
Note that such design naturally respects also the properties of the state noise $\bfw_k$ and each grid point is associated with the same neighbourhood $\bfDelta_{k+1}$.

\subsubsection{Grid for Interpolation of Filtering PMD:}
To enable the Lagrangian solution to the GbF prediction, the predictive grid has not only to be equidistant (which is ensured by its design) but it must also be unequivocally related to the grid of the filtering PMD. To ensure the latter property, let the new grid $\bfXi_{k+1}$ be back-propagated via an inverse state dynamics according to%to get the grid $\bfXi_{k}^{\mathrm{BP}}$ on which the filtering PMD $\bar{p}(\bfx_k|\bfz^{k};\bfXi_k)$ is interpolated. As a consequence, we have the filtering PMD $\bar{p}(\bfx_k|\bfz^{k};\bfXi_{k}^{\mathrm{BP}})$. %The actual construction of the grid for interpolation depends on the properties of the state dynamics, particularly whether the function $\bff_k(\bfx_k)$ is invertible $\forall \bfx_k$ or not.
%Considering the invertible state dynamics $\bff_k$, the back-propagated grid can directly be constructed as
\begin{align}
    \bfxi_{k}^{\mathrm{BP},(i)}=\bff_k^{-1}(\bfxi_{k+1}^{(i)}),\forall i.\label{eq:xi_interp_inv}
\end{align}
%where $\bff_k^{-1}$ denotes the inverse of the function $\bff_k$ with a property $\bff_k^{-1}(\bff_k(\bfx_k))=\bfx_k, \forall \bfx_k$. 
Generally, the grid $\bfXi_{k}^{\mathrm{BP}} = \big\{\bfxi_{k}^{\mathrm{BP},(i)}\big\}_{i=1}^N$ is not structured, i.e., it is not equidistant and each point is associated with its own neighbourhood $\bfDelta_k^{\mathrm{BP},(i)}$ on which the probability value (i.e., the weight) is assumed to be constant. The neighbourhood of the $i$-th point $\bfxi_{k}^{\mathrm{BP},(i)}$ \eqref{eq:xi_interp_inv} can be approximately (yet efficiently) calculated using the Taylor expansion of the state dynamics at each back-propagated grid point as
\begin{align}
    \bfx_{k+1}^{\mathrm{adv}}\approx\bff_k\left(\bfxi_{k}^{\mathrm{BP},(i)}\right)&+\bfJ_{\bff_k}\left(\bfxi_{k}^{\mathrm{BP},(i)}\right)\left(\bfx_k-\bfxi_{k}^{\mathrm{BP},(i)}\right)\nonumber\\
    \bfx_{k+1}^{\mathrm{adv}}-\bfxi_{k+1}^{(i)}&\approx\bfJ_{\bff_k}\left(\bfxi_{k}^{\mathrm{BP},(i)}\right) \left(\bfx_k-\bfxi_{k}^{\mathrm{BP},(i)}\right). %\label{eq:linTE}
\end{align}
where $\bfJ_{\bff_k}(\bbfx)=\tfrac{\partial \bff_k(\bfx_k)}{\partial \bfx_k}|_{\bfx_k=\bbfx}$ is the Jacobian\footnote{Matrix of first-order differences can be used instead of the Jacobian (leading to Stirling's interpolation), if derivative is not to be calculated.  Higher-order interpolation can be used in case of highly nonlinear dynamics.} of $\bff_k$ evaluated at $\bbfx$. Due to the definition of the selection function $S$ in \eqref{eq:pmd} and the \textit{locally} valid linearisation of $\bff_k$ at a grid point $\bfxi_{k}^{\mathrm{BP},(i)}$, % we can assume that if $\bfx_{k+1}^{\mathrm{adv}}\in S(\bfx_{k+1};\bfxi_{k+1}^{(i)},\bfDelta_{k+1})$, then $\bfx_{k}\in S(\bfx_k;\bfxi_{k}^{\mathrm{BP},(i)},\bfDelta_{k}^{(i)})$. As such, \eqref{eq:linTE} can be written as
we can write
%\begin{align}
    $\bfDelta_{k+1}\approx\bfJ_{\bff_k}(\bfxi_{k}^{\mathrm{BP}(i)})\bfDelta_{k}^{(i)}$. % \label{eq:linTE}
%\end{align}
That means the approximate neighbourhood of $i$-th back-propagated point reads
\begin{align}
    \bfDelta_k^{\mathrm{BP},(i)}\approx\bfJ_{\bff_k^{-1}}(\bfxi_{k+1}^{(i)})\bfDelta_{k+1},
\end{align}
where the equality $\bfJ_{\bff_k^{-1}}(\bfxi_{k+1}^{(i)})=\left(\bfJ_{\bff_k}(\bfxi_k^{\mathrm{BP},(i)})\right)^{-1}$ was used. It should be noted that if $\bfDelta_{k+1}$ is a rectangular neighbourhood of a grid point $\bfxi_{k+1}^{(i)}$, the neighbourhood $\bfDelta_k^{\mathrm{BP},(i)}$ defines an $n_x$-dimensional paralleltope\footnote{The parallelotope is the image of a hyper-rectangle under a linear transformation (plus possibly a translation).}.

\subsubsection{Filtering PMD Interpolation:}
Having the back-propagated grid $\bfXi_k^{\mathrm{BP}}$ given by \eqref{eq:xi_interp_inv}, the filtering PMD $\bar{p}(\bfx_k|\bfz^{k};\bfXi_k)$ can be interpolated on the grid leading to the interpolated filtering PMD $\bar{p}(\bfx_k|\bfz^{k};\bfXi_k^{\mathrm{BP}})$.
That is, the weights ${P}_{k|k}(\bfxi_k^{(i)})$ on $\bfXi_k$ are interpolated to yield the weights ${P}_{k|k}(\bfxi_k^{\mathrm{BP}(i)})$ on $\bfXi_{k}^{\mathrm{BP}}$.
%Computational complexity of the grid-back propagation and the filtering PMD interpolation is $\calO(N)$.

\subsubsection{Advected Predictive PMD Calculation:}
The PMD of the advected predicted state $\bfx_{k+1}^{\mathrm{adv}}$ can be constructed simply as
\begin{align}
    \bar{p}(\bfx_{k+1}^{\mathrm{adv}}|\bfz^k;\bfXi_{k+1})%\nonumber\\
    \triangleq\sum_{i=1}^N{P}_{k+1|k}^{\mathrm{adv}}(\bfxi_{k+1}^{(i)})S(\bfx_{k+1};\bfxi_{k+1}^{(i)},\bfDelta_{k+1}),\label{eq:pmd_pred_adv}
\end{align}
where 
\begin{align}
    {P}_{k+1|k}^{\mathrm{adv}}(\bfxi_{k+1}^{(i)}) = {P}_{k|k}(\bfxi_{k}^{\mathrm{BP},(i)})\tfrac{\delta_k^{\mathrm{BP},(i)}}{\delta_{k+1}}.\label{eq:pmd_filt_norm}
\end{align}
% In this view,
It can be seen that the grid is synchronously moved with the state dynamics\footnote{
    In the fluid dynamics area, similar grid motion is employed in Lagrangian-based methods, thus the name ``Lagrangian'' GbF.
} as $\bfxi_{k+1}^{(i)} = \bff_k(\bfxi_{k}^{\mathrm{BP},(i)})$ due to \eqref{eq:xi_interp_inv}.
% In the fluid dynamics setting, similar grid motion is employed in Lagrangian-based methods, thus the name Lagrangian GbF. % leading to \textcolor{red}{???}.
% In the state estimation setting, the above design leads the grid points being equidistantly spaced at time $k+1$, which is the key enabler for efficient diffusion solution;
Besides that the grid $\bfXi_{k+1}$ is equidistant, which enables the following efficient \textit{diffusion} solution. 
%The key advantage of the above design is that the grid points are equidistantly spaced at time $k+1$, which enables efficient diffusion solution.
As the computational complexity of grid back-propagation and gridded interpolation is $\calO(N)$, the complexity of the entire advection is also $\calO(N)$.%which enables efficient solution of the prediction step by the convolution.

\subsection{Diffusion Solution}
What is left to be solved is the diffusion part corresponding to model \eqref{eq:adv_dif}, which describes a sum of two random variables (RVs) with known densities.  The density of the sum of the RVs is thus equal to the convolution of the summands' PDFs (i.e.,  PMDs). The convolution can be efficiently solved by the DFT; thus, the diffusion solution for the predictive PMD weight update becomes
\begin{align}
	{{P}}_{k+1|k}=\mathcal{F}^{-1}\left(  \mathcal{F}({W}_k) \odot \mathcal{F}({{{P}}_{k+1|k}^{\mathrm{adv}}})\right), \label{eq:tpmconvDFT}%\widetild{\mathcal{P}}_{k+1}^{(:)} = \mathcal{F}^{-1}\left(  \mathcal{F}(\widetilde{\bfT^{(m,:)}_k}) \odot \mathcal{F}(\widetilde{ \mathcal{P}}_{{k}}^{(:)})\right), 
\end{align}
where $\mathcal{F}(\cdot)$ denotes the DFT, $\mathcal{F}^{-1}(\cdot)$ is the inverse DFT, and the tensor ${W}_k= p(\bfxi_{k+1}^\mathrm{mean}|\bfXi_{k+1}) \in \mathbb{R}^{N_1 \times N_2 \times ... \times N_{n_x}}$ is the state transition probability evaluated at the old grid points with $\bfxi_{k+1}^\mathrm{mean}$ being the mean over all points in $\bfXi_{k+1}$. If $N_i$ is set odd $\forall i$, then $\bfxi_{k+1}^\mathrm{mean}$ is one of the grid points in $\bfXi_{k+1}$, which simplifies the calculations. Due to the DFT, the computational complexity of the diffusion is $\calO(N\log N)$.

\subsection{Algorithm Summary and Prediction Illustration}
\begin{figure*}[]
	\centering
	\includegraphics[width=1\linewidth]{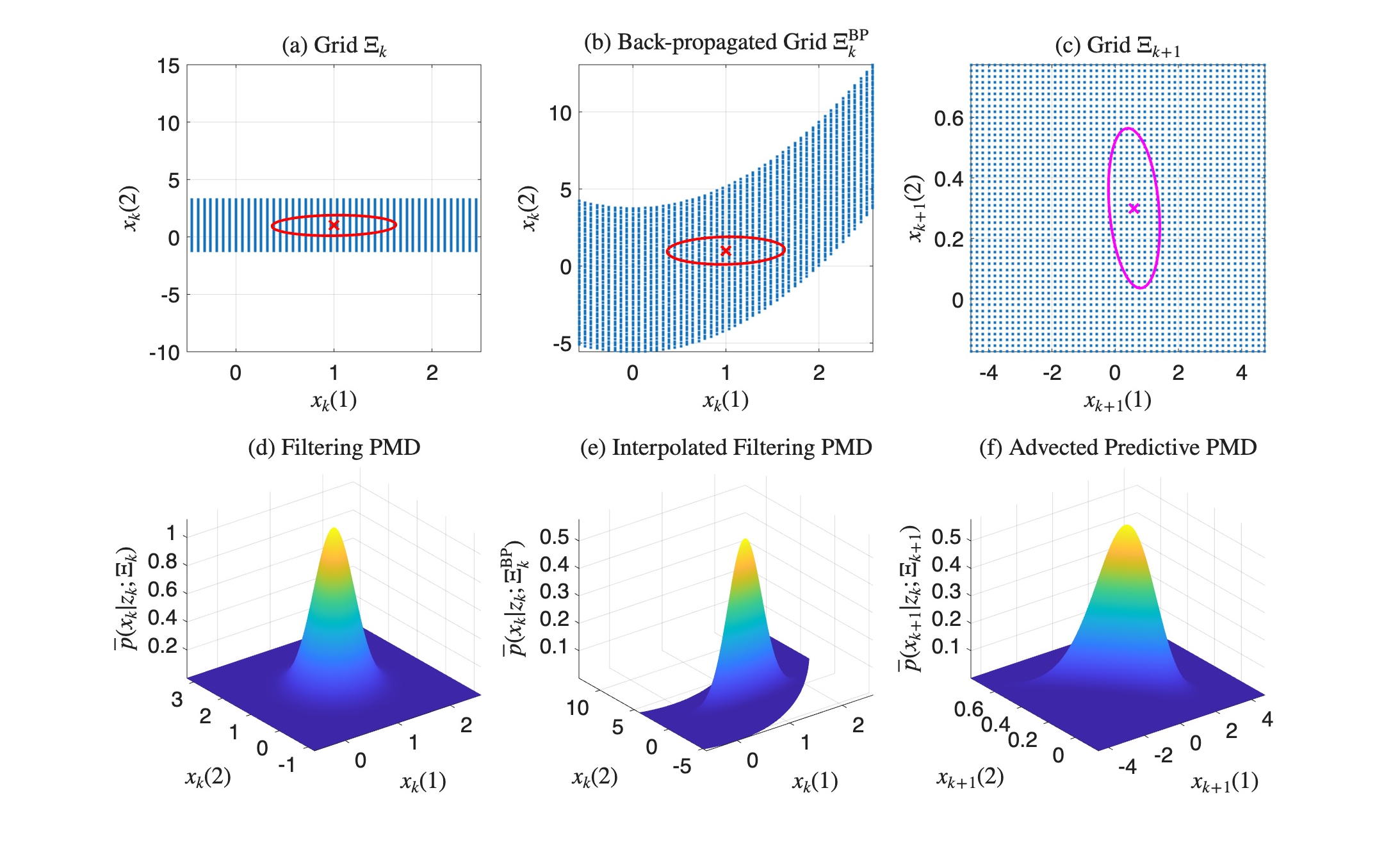}\vspace*{-2mm}
	\caption{Illustration of steps of LGbF prediction using the Hénon model.}
	\label{fig:LGbFpred}\vspace*{-0mm}
\end{figure*}

%\subsubsection{Time Update Algorithm Summary:}
{The proposed LGbF follows the structure of any Bayesian filter given, in particular, by the initialisation, the measurement update (i.e., filtering step), and the time update (i.e., prediction step).}

{\subsubsection{Initialisation:} The initialisation is identical with the initialisation of the EGbF given in Section \ref{sec:EGbF}. The initialisation provide the initial PMF $ \bar{p}(\bfx_0|\bfz^{-1};\bfXi_0)$ according to \eqref{eq:pmd}.}

{\subsubsection{Measurement Update:} The filtering step is analogous to the one of the  EGbF given in Section \ref{sec:EGbF}. Having the predictive (or initial) PMD $ \bar{p}(\bfx_k|\bfz^{k-1};\bfXi_k)$, the weights of the filtering PMD $\bar{p}(\bfx_k|\bfz^{k};\bfXi_k)$ are calculated $\forall i$ according to \eqref{eq:pmf_filtrace_a}.}

%the EGbF given in Section \ref{sec:EGbF}. In particular, the \textit{initialisation} and \textit{filtering} remains identical.
% The only change is in the \textit{prediction}, which is can be summarised in the following steps.
\subsubsection{Time Update (with Illustration):} The advancements made to the \textit{prediction} are summarised in the following steps and illustrated in Fig. \ref{fig:LGbFpred} using the Hénon map model defined later in Section \ref{sec:Henon};

\begin{enumerate}[label=(\roman*)]
    \item Given the filtering PMD $\bar{p}(\bfx_k|\bfz^{k};\bfXi_k)$  with weights $P_{k|k}(\bfxi_k^{(i)})$~\eqref{eq:pmf_filtrace_a} evaluated on rectangular grid $\bfXi_k$, calculate the filtering mean $\hbfx_{k|k}$~\eqref{eq:mean} and covariance matrix $\bfC_{k|k}$~\eqref{eq:cov}.
    % The grid $\bfXi_k$ with associated filtering PMD weights $P_{k|k}(\bfxi_k^{(i)})$ can be seen in Fig. \ref{fig:LGbFpred}(a) and \ref{fig:LGbFpred}(d), respectively. The filtering mean and covariance matrix are plotted in red in Figure \ref{fig:LGbFpred}(a) as well.
    Fig.~\ref{fig:LGbFpred}(a) shows the grid $\bfXi_k$ in blue along with the corresponding mean (red cross) and covariance matrix depicted as an ellipse (red), while
    Fig.~\ref{fig:LGbFpred}(d) shows the associated filtering PMD weights $P_{k|k}(\bfxi_k^{(i)})$.
    \item Calculate the predictive mean $\hbfx_{k+1|k}$ and covariance matrix $\bfC_{k+1|k}$ using the prediction step of a chosen local filter. The predictive mean and covariance matrix are shown in magenta in Fig.~\ref{fig:LGbFpred}(c).
    \item Based on the predicted moments {and scaling factor $\kappa$}, design the rectangular grid $\bfXi_{k+1}$, which can be seen in Fig.~\ref{fig:LGbFpred}(c) in blue.
    \item  Using the inverse dynamics $\bff_k^{-1}$, calculate the back\discretionary{-}{-}{-}propagated grid $\bfXi_{k}^{\mathrm{BP}}$ according to~\eqref{eq:xi_interp_inv}. The back\discretionary{-}{-}{-}propagated grid is visualised in Fig.~\ref{fig:LGbFpred}(b).
    \item Interpolate the filtering PMD $\bar{p}(\bfx_k|\bfz^{k};\bfXi_k)$ on the back-propagated grid $\bfXi_{k}^{\mathrm{BP}}$, i.e., calculate $\bar{p}(\bfx_k|\bfz^{k};\bfXi_k^{\mathrm{BP}})$. The filtering PMD on $\bfXi_{k}^{\mathrm{BP}}$ is illustrated in Fig.~\ref{fig:LGbFpred}(e).
    \item Calculate the weights ${P}_{k+1|k}^{\mathrm{adv}}(\bfxi_{k+1}^{(i)})$ of the advected predictive PMD according to \eqref{eq:pmd_filt_norm} on the grid $\bfXi_{k+1}$.
    \item Using the state noise PDF $p(\bfw_k)$, calculate weights ${{P}}_{k+1|k}$ of the the predictive PMD $\bar{p}(\bfx_{k+1}|\bfz^{k};\bfXi_{k+1})$ using DFT as defined in \eqref{eq:tpmconvDFT}. The resulting predictive PMD can be found Fig.~\ref{fig:LGbFpred}(f).
\end{enumerate}

{\subsection{Integration Error and Implementation Notes}}
{Having derived the LGbF for the nonlinear state-space model with invertible dynamics, we provide a few notes on implementation and integration error.}
{\subsubsection{Integration Error Bounds:} The introduced EGbF and LGbF are both based on the midpoint numerical integration rule, which is the main source of the error in the solution to the BRRs \eqref{eq:filt}, \eqref{eq:pred}. Assuming the equidistantly spaced grids, as it is in the EGbF or the LGbF for the linear state dynamics, the midpoint rule global accuracy is of the second order $\calO(\delta_k^2)$ \citep{StBu:92}. The proposed LGbF has, however, the non-equidistantly spaced grid $\bfXi_{k}^{\mathrm{BP}}$ due to the back-propagation through the nonlinear function \eqref{eq:xi_interp_inv}. In this case, thus, the upper bound of the global accuracy becomes $\calO((\delta_k^{\mathrm{max}})^2)$ with $\delta_k^{\mathrm{max}}=\max_i\delta_k^{(i)}$. Note that second order accuracy means that doubling the number grid points $N$ leads roughly to a quartered numerical integration error.}

\subsubsection{Computationally Efficient Implementation:}
Due to the diffusion step, the overall complexity of the LGbF is $\calO(N\log N)$, which represents a significant reduction w.r.t. the EGbF complexity of $\calO(N^2)$. Nevertheless, the resulting computational time depends on particular implementation and several suggestions for efficient implementation are given below. 
\begin{itemize}
    \item Efficient interpolation of scattered grid $\bfXi_{k}^{\mathrm{BP}}$ in Step (v) can be done using, e.g., the MATLAB\textregistered\ routine \verb|griddedInterpolant|.
    \item If $\delta_k^{\mathrm{BP},(i)}$ is independent of the back-propagated grid point, then the weight ${P}_{k+1|k}^{\mathrm{adv}}(\bfxi_{k+1}^{(i)})$ need not be normalised \textit{point-by-point} by the ratio of the neighbourhood volumes $\tfrac{\delta_k^{\mathrm{BP},(i)}}{\delta_{k+1}}$ \eqref{eq:pmd_filt_norm}. Instead, the whole predictive PMD can be simply normalised as it is done after any operation.
\end{itemize}

{\subsubsection{Models with Non-invertible Dynamics:} If the dynamics in \eqref{eq:asx} is not invertible, we can either
\begin{itemize}
    \item Use the two-grid-approach proposed in \citep{DuMaSt:23} with carefully chosen fine grid, which can be seen as a brute-force solution for the filtering PMD propagation via $\bff_k$ in the prediction step,
    \item Linearise $\bff_k$ and then use the LGbF for linear dynamics \citep{MaDuSt:25}. 
\end{itemize}}

\section{Numerical Illustration and Implementation}
Two numerical examples were implemented to verify the performance of the proposed LGbF w.r.t. the EGbF and the PF using the open MATLAB\textregistered{} code\footnote{The main simulation is executed via \textit{main.m}, where a model can be selected. The state-space model configured in \textit{initModel.m}, allowing both the state and the measurement to have arbitrary dimensions. }.%The repository is \url{https://github.com/pesslovany/Matlab-LagrangianPMF}.}. 

 %Illustration of complexities for different approaches can be seen in Table \ref{tab:res}. The accuracy and consistency of LGbE and EGbE is qualitatively the same. Comparison of GbFs with other filters were done in various other papers \citep{MaDuSt:25,AnHa:06,OrSkToGu:10}, and due to the space constraint is not presented here.  
\subsection{State Estimation of Chaotic System: Hénon Map (2D)}\label{sec:Henon}
In the first example, the dynamical system with chaotic behaviour is estimated. Namely, the Hénon map with $n_x=2$ is considered, %which is one of the most studied examples of dynamical systems that exhibit chaotic behaviour. For the Hénon map, 
where the dynamics~\eqref{eq:asx} reads \citep{He:76}
\begin{align}
    \bff_k(\bfx_k)=\begin{bmatrix}
                    1 - a \cdot (\bfx_k(1))^2 + \bfx_k(2) \\
                    b\cdot \bfx_k(1)
                    \end{bmatrix},
\end{align}
where $\bfx_k(1)$ represents the main coordinate that undergoes quadratic stretching and folding, $\bfx_k(2)$ acts as the memory variable, the parameter $a=1.4$ is the ``chaotic intensity'' parameter controlling the strength of the nonlinearity, % and therefore the degree of stretching and folding in the map, 
and the parameter $b=0.3$ %determines the amount of contraction along one direction of the state space and therefore, it 
acts as a %dissipation or 
memory coefficient. The measurement equation~\eqref{eq:asz} is assumed to be linear with
\begin{align}
    h_k(\bfx_k)= \big[1\ \ 0\big] \, \bfx_k,
\end{align}
i.e., $n_z=1$. The state and measurement noises are assumed to be white and zero mean with the covariance matrices
\begin{align}
\bfQ=\cov[\bfw_k]=\left[\begin{smallmatrix}
    10^{-3} & 0\\ 0 & 10^{-5}
\end{smallmatrix}\right], \ R=\var[v_k]=0.01.
\end{align}
Design of the LGbF requires inversion and Jacobian of the state dynamics, which are given by
\begin{align}
    \bff_k^{-1}(\bfx_{k+1}^\mathrm{adv})&=\left[\begin{smallmatrix}
b^{-1}\cdot{\bfx_{k+1}^\mathrm{adv}(2)} \\
\bfx_{k+1}^\mathrm{adv}(1) - 1 + a\cdot b^{-2}\cdot\left({\bfx_{k+1}^\mathrm{adv}(2)}\right)^{\!2}
\end{smallmatrix}\right], \\
\bfJ_{\bff_k}(\bfx_k) &=
\left[\begin{smallmatrix}
-2a\cdot\bfx_k(1) & 1 \\
b & 0
\end{smallmatrix}\right].
\end{align}

% \begin{align}
%     \bff_k^{-1}(\bfx_{k+1}^\mathrm{adv})&=\left[\begin{smallmatrix}
% \tfrac{\bfx_{k+1}^\mathrm{adv}(2)}{b} \\
% \bfx_{k+1}^\mathrm{adv}(1) - 1 + a\!\left(\tfrac{\bfx_{k+1}^\mathrm{adv}(2)}{b}\right)^{\!2}
% \end{smallmatrix}\right], \\
% \bfJ_{\bff_k}(\bfx_k) &=
% \left[\begin{smallmatrix}
% -2a\,\bfx_k(1) & 1 \\
% b & 0
% \end{smallmatrix}\right].
% \end{align}

Several local, grid-based, and particle filters have been implemented and are available in the GitHub repository. For this paper four filters were selected for performance illustration, namely, the \textit{developed LGbF}, \textit{standard EGbF}, recent \textit{Gaussian mixture-based point-mass filter} (GM-PMF) \citep{GiPoZa:24}, and the \textit{PF} with systematic resampling \citep{DoFrGo:01}. The LGbF, EGbF, and PF were set to have the same number of grid points or particles, namely $N=31^2=961$. The performance was evaluated over 100 MC simulations with $k=0,1,\ldots,10$ according to three criteria:
\begin{itemize}
    \item Root-mean-square error (RMSE) assessing the accuracy (the lower, the better),
    \item Averaged normalised estimate error squared (ANEES) assessing the consistency (closer to one, the better),
    \item Computational time.
\end{itemize}
Mathematical definition of the first two criteria can be found e.g., in \citep{MaDuSt:25}.

\begin{table}%[h]
    \centering
    \caption{2D Example: Performance metrics.}\vspace*{-2mm}
\begin{tabular}{l|ccc}
Technique & RMSE & ANEES  & Time [sec] \\ 
\hline 
\textbf{LGbF} & 0.052 & 0.99 & $4\times10^{-4}$ \\ 
EGbF & 0.052 & 1.22 & $1\times10^{-2}$ \\ 
GM-PMF & 0.052 & 0.94 &  $2\times10^{-2}$ \\ 
PF & 0.052 & 1.1 & $1\times10^{-4}$ \\ 
\hline 
\end{tabular}
    \label{tab:2d}
\end{table}

The results are summarised in Table \ref{tab:2d}. The results indicate that all the considered filters provide very similar performance. This is true especially for the RMSE, that was nearly the same for all the filters. While considering the ANEES, the proposed LGbF provided the most consistent estimates, although it has slightly higher computational complexity than the PF. Nevertheless, the proposed LGbF is two orders of magnitude faster than both the standard EGbF and the recent GM-PMF. %that the proposed LGbF is about %Note that the standard EGbF provides the same results as the LGbF but with two orders higher computational complexity.
{It is worth noting that the two-grid-based implementation of the LGbF \citep{DuMaSt:23} reaches, in this case,  the same RMSE and ANEES as the proposed LGbF but with more than twice higher computational complexity.}

\subsection{Tracking: Turn Model with Unknown Rate (5D)}
In the second example, we consider the object tracking using the five dimensional coordinated turn model with unknown rate with radar measurements in the form of range and bearing \citep{BaLiKi:01}. The nonlinear functions of the state-space model \eqref{eq:asx}, \eqref{eq:asz} with $n_x=5, n_z=2$ are as follows
\begin{align}
    \bff_k(\bfx_k)\! &=\!\!
\left[\begin{smallmatrix}
\bfx_k(1) + \tfrac{\sin(\bfx_k(5))}{\bfx_k(5)} \bfx_k(2) - \tfrac{1 - \cos(\bfx_k(5))}{\bfx_k(5)} \bfx_k(4) \\
\cos(\bfx_k(5))\, \bfx_k(2) - \sin(\bfx_k(5))\, \bfx_k(4) \\
\bfx_k(3) + \tfrac{1 - \cos(\bfx_k(5))}{\bfx_k(5)} \bfx_k(2) + \tfrac{\sin(\bfx_k(5))}{\bfx_k(5)} \bfx_k(4) \\
\sin(\bfx_k(5))\, \bfx_k(2) + \cos(\bfx_k(5))\, \bfx_k(4) \\
\bfx_k(5)
\end{smallmatrix}\right]\!\!,\label{eq:5Df}\\
\bfh_k(\bfx_k)\!&=\!\!\left[\begin{smallmatrix}
    \operatorname{atan2}(\bfx_k(3),\, \bfx_k(1)), &
    \sqrt{(\bfx_k(1))^2 + (\bfx_k(3))^2}
\end{smallmatrix}\right]^T, 
% \bfh_k(\bfx_k)\!&=\!\!\left[\begin{smallmatrix}
%     \operatorname{atan2}(\bfx_k(3),\, \bfx_k(1)) \\[0.5em]
%     \sqrt{(\bfx_k(1))^2 + (\bfx_k(3))^2}
% \end{smallmatrix}\right], 
\end{align}
where $\bfx_k(1),\ \bfx_k(3)$ stand for the horizontal (north-east) position of the object, $\bfx_k(2),\ \bfx_k(4)$ for the velocity, and $\bfx_k(5)$ for the turn rate. The first component of the measurement is the object bearing and the second component is the range (both w.r.t. radar located at coordinate system centre).

The inverse, needed for the LGbF design, is
% \begin{align*}
%     \bff_k^{-1}(\bfx_{k+1}^\mathrm{adv}) &=
% \begin{smallmatrix}
% \bfx_{k+1}^\mathrm{adv}(1) - \tfrac{\sin \bfx_{k+1}^\mathrm{adv}(5)}{\bfx_{k+1}^\mathrm{adv}(5)} (\cos \bfx_{k+1}^\mathrm{adv}(5)\, \bfx_{k+1}^\mathrm{adv}(2) + \sin \bfx_{k+1}^\mathrm{adv}(5)\, \bfx_{k+1}^\mathrm{adv}(4))
%       + \tfrac{1 - \cos \bfx_{k+1}^\mathrm{adv}(5)}{\bfx_{k+1}^\mathrm{adv}(5)} (-\sin \bfx_{k+1}^\mathrm{adv}(5)\, \bfx_{k+1}^\mathrm{adv}(2) + \cos \bfx_{k+1}^\mathrm{adv}(5)\, \bfx_{k+1}^\mathrm{adv}(4)) \\[0.7em]
% \cos \bfx_{k+1}^\mathrm{adv}(5)\, \bfx_{k+1}^\mathrm{adv}(2) + \sin \bfx_{k+1}^\mathrm{adv}(5)\, \bfx_{k+1}^\mathrm{adv}(4) \\[0.7em]
% \bfx_{k+1}^\mathrm{adv}(3) - \tfrac{1 - \cos \bfx_{k+1}^\mathrm{adv}(5)}{\bfx_{k+1}^\mathrm{adv}(5)} (\cos \bfx_{k+1}^\mathrm{adv}(5)\, \bfx_{k+1}^\mathrm{adv}(2) + \sin \bfx_{k+1}^\mathrm{adv}(5)\, \bfx_{k+1}^\mathrm{adv}(4))
%       - \tfrac{\sin \bfx_{k+1}^\mathrm{adv}(5)}{\bfx_{k+1}^\mathrm{adv}(5)} (-\sin \bfx_{k+1}^\mathrm{adv}(5)\, \bfx_{k+1}^\mathrm{adv}(2) + \cos \bfx_{k+1}^\mathrm{adv}(5)\, \bfx_{k+1}^\mathrm{adv}(4)) \\[0.7em]
% -\sin \bfx_{k+1}^\mathrm{adv}(5)\, \bfx_{k+1}^\mathrm{adv}(2) + \cos \bfx_{k+1}^\mathrm{adv}(5)\, \bfx_{k+1}^\mathrm{adv}(4) \\[0.7em]
% \bfx_{k+1}^\mathrm{adv}(5)
% \end{smallmatrix},
% \end{align*}
\begin{align*}
    \bff_k^{-1}(\bfx_{k+1}^\mathrm{adv}) &=\left[
\begin{smallmatrix}
\bfx_{k+1}^\mathrm{adv}(1) - \tfrac{\sin(\bfx_{k+1}^\mathrm{adv}(5))}{\bfx_{k+1}^\mathrm{adv}(5)} \eta_1
      + \tfrac{1 - \cos(\bfx_{k+1}^\mathrm{adv}(5))}{\bfx_{k+1}^\mathrm{adv}(5)} \eta_2 \\
\eta_1 \\
\bfx_{k+1}^\mathrm{adv}(3) - \tfrac{1 - \cos(\bfx_{k+1}^\mathrm{adv}(5))}{\bfx_{k+1}^\mathrm{adv}(5)} \eta_1
      - \tfrac{\sin(\bfx_{k+1}^\mathrm{adv}(5))}{\bfx_{k+1}^\mathrm{adv}(5)} \eta_2 \\
\eta_2 \\
\bfx_{k+1}^\mathrm{adv}(5)
\end{smallmatrix}\right]
\end{align*}
with $\eta_1=(\cos(\bfx_{k+1}^\mathrm{adv}(5))\, \bfx_{k+1}^\mathrm{adv}(2) + \sin(\bfx_{k+1}^\mathrm{adv}(5))\, \bfx_{k+1}^\mathrm{adv}(4))$ and $\eta_2=(-\sin(\bfx_{k+1}^\mathrm{adv}(5))\, \bfx_{k+1}^\mathrm{adv}(2) + \cos(\bfx_{k+1}^\mathrm{adv}(5))\, \bfx_{k+1}^\mathrm{adv}(4))$. The Jacobian of \eqref{eq:5Df} can be calculated either analytically or numerically using the difference. The standard setting of the model together with the noise properties can be found in \citep{BaLiKi:01} or the source code at repository.

\begin{figure}[]
	\centering
	\includegraphics[width=0.8\linewidth]{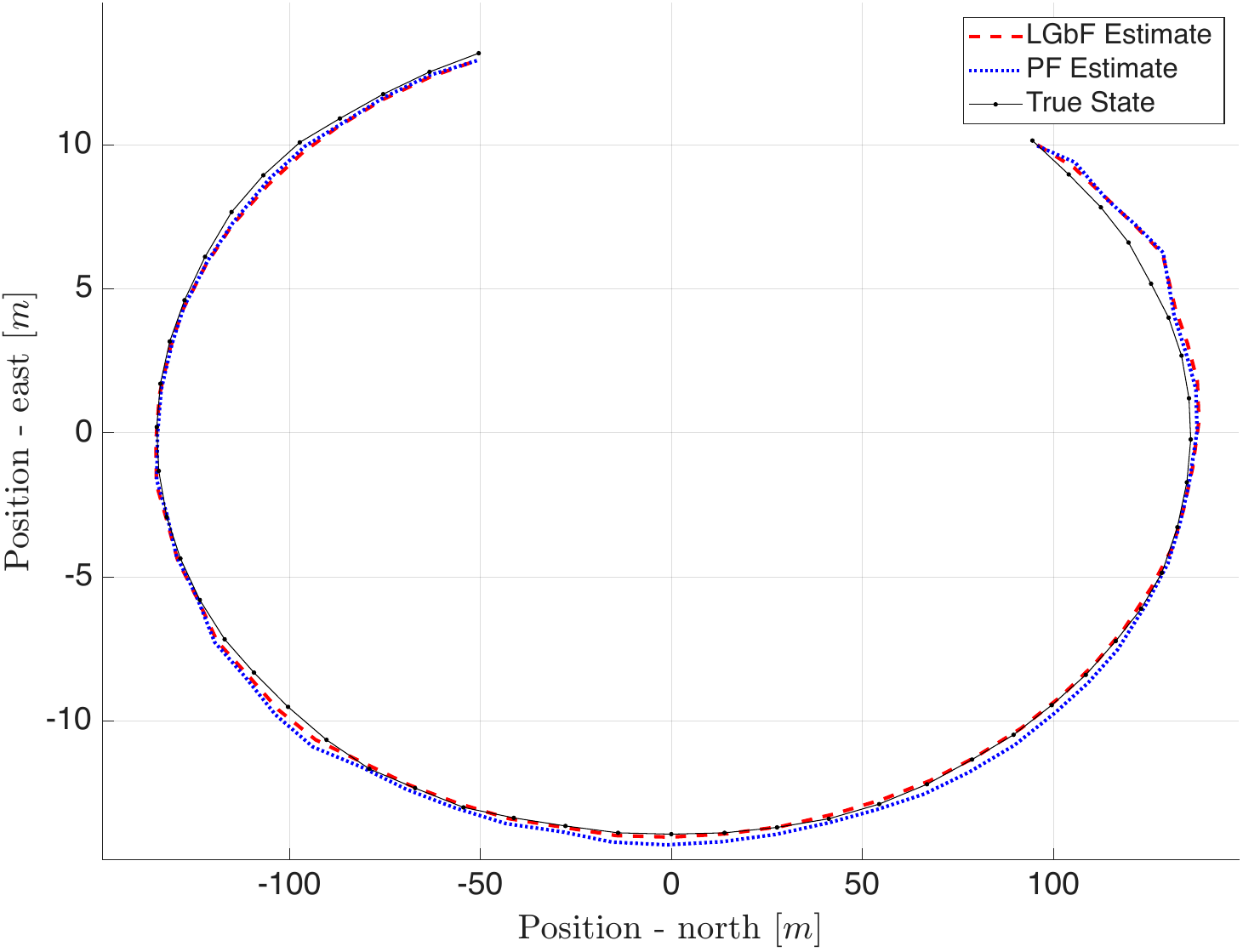}\vspace*{-2mm}
	\caption{Performance illustration of tracking using 5D model.}
	\label{fig:tracking}\vspace*{-0mm}
\end{figure}

Similarly to the previous example, several filters were implemented, but we present the results for two of them only, namely for the LGbF and the PF with $N=21^5$.
Note that the evaluation of the the standard EGbF computational complexity is about hundreds of hours per one time update for the considered large $N$ due to the demanding evaluation of~\eqref{eq:pred}.
The true and estimated trajectories are illustrated in Fig.~\ref{fig:tracking}.

Both filters provide consistent and similar results. The LGbF is more demanding than the PF (seconds vs. tenths of seconds). However, the LGbF preserves the advantage of being deterministic, which is important for safety-critical applications (including the tracking and surveillance systems).

% \begin{align}
%     J_f(x,u) &=
% \frac{\partial f(x,u)}{\partial x} =
% \begin{bmatrix}
% 1 & \dfrac{\sin x_5}{x_5} & 0 & -\dfrac{1 - \cos x_5}{x_5} & \star \\[0.7em]
% 0 & \cos x_5 & 0 & -\sin x_5 & -x_2 \sin x_5 - x_4 \cos x_5 \\[0.7em]
% 0 & \dfrac{1 - \cos x_5}{x_5} & 1 & \dfrac{\sin x_5}{x_5} & \star \\[0.7em]
% 0 & \sin x_5 & 0 & \cos x_5 & x_2 \cos x_5 - x_4 \sin x_5 \\[0.7em]
% 0 & 0 & 0 & 0 & 1
% \end{bmatrix},
% \end{align}

%\subsection{Simulation Results}

%For completeness, this subsection presents results for a 2D model. The parameters and model setup are provided in the accompanying MATLAB\textregistered\ implementation. 

%Table \ref{tab:res} shows the root mean square error (RMSE) for filtering and smoothing estimates, as well as the time required to perform one step of estimation and smoothing. The top section presents results when transition probabilities ($\bfT_k$ resp. $\widetilde{\bfW}_k$) are saved during the forward run of the estimator, while the bottom section shows results when the transition probabilities are re-calculated during smoothing.

%As expected, the smoothing estimates are more accurate than the filtering estimates. The RMSE difference between the LGbF and standard GbF methods arises from differences in grid design. When transition probabilities are saved, the standard GbF requires a substantially more memory than LGbF. Conversely, when the values are not saved, the standard GbF is slower by two orders of magnitude.

\section{Conclusion}
The paper dealt with the state estimation of nonlinear and non-Gaussian stochastic dynamic systems using the grid-based filters. The emphasis was laid on the development of Lagrangian grid-based filter providing consistent and accurate estimates for models with nonlinearities not only in the measurement but also in the state equation.  %in both measurement and state equations. % not only the measurement but also the state equation. 
Compared to the standard Eulerian implementation, the LGbF reduces the computational complexity from quadratic to quasilinear (or log-linear), making the algorithm an appealing option for higher dimensional models. % with state dimension up to seven. 
LGbF performance was illustrated %in two studies 
using the open codes.% that are available in the repository. %, and the proposed filter was shown to be viable for models with state dimension up to seven.

% \section*{DECLARATION OF GENERATIVE AI AND AI-ASSISTED TECHNOLOGIES IN THE WRITING PROCESS}
% During the preparation of this work the author(s) used [NAME TOOL / SERVICE] in order to [REASON]. After using this tool/service, the author(s) reviewed and edited the content as needed and take(s) full responsibility for the content of the publication.

%\bibliography{literatura}             % bib file to produce the bibliography

\begin{thebibliography}{15}
\providecommand{\natexlab}[1]{#1}
\providecommand{\url}[1]{\texttt{#1}}
\providecommand{\urlprefix}{URL }
\expandafter\ifx\csname urlstyle\endcsname\relax
  \providecommand{\doi}[1]{doi:\discretionary{}{}{}#1}\else
  \providecommand{\doi}{doi:\discretionary{}{}{}\begingroup \urlstyle{rm}\Url}\fi

\bibitem[{Anderson and Moore(1979)}]{AnMo:79}
Anderson, B.D.O. and Moore, J.B. (1979).
\newblock \emph{Optimal Filtering}.
\newblock Prentice Hall, New Jersey.

\bibitem[{Ånonsen and Hallingstad(2006)}]{AnHa:06}
Ånonsen, K.B. and Hallingstad, O. (2006).
\newblock Terrain aided underwater navigation using point mass and particle filters.
\newblock In \emph{2006 IEEE/ION Position, Loc., and Nav. Symp.}, 1027--1035.

\bibitem[{Ånonsen and Hagen(2010)}]{AnHa:10}
Ånonsen, K.B. and Hagen, O.K. (2010).
\newblock An analysis of real-time terrain aided navigation results from a {HUGIN AUV}.
\newblock In \emph{2010 Marine Technology Society (MTS) and the Oceanic Engineering Society of the IEEE Conference}.

\bibitem[{Bar-Shalom et~al.(2001)Bar-Shalom, Li, and Kirubarajan}]{BaLiKi:01}
Bar-Shalom, Y., Li, X.R., and Kirubarajan, T. (2001).
\newblock \emph{Estimation with Applications to Tracking and Navigation: Theory Algorithms and Software}.
\newblock John Wiley \& Sons.

\bibitem[{Bergman(1999)}]{Be:99}
Bergman, N. (1999).
\newblock \emph{Recursive Bayesian Estimation: Navigation and Tracking Applications}.
\newblock Ph.D. thesis, Link\"{o}ping University, Sweden.

\bibitem[{Bucy and Senne(1971)}]{BuSe:71}
Bucy, R.S. and Senne, K.D. (1971).
\newblock Digital synthesis of non-linear filters.
\newblock \emph{Automatica}, 7, 287--298.

\bibitem[{Choe and Park(2021)}]{ChGo:21}
Choe, Y. and Park, C.G. (2021).
\newblock Point-mass filtering with boundary flow and its application to terrain referenced navigation.
\newblock \emph{IEEE Trans. on AES}, 57(6), 3600--3613.

\bibitem[{Doucet et~al.(2001)Doucet, De~Freitas, and Gordon}]{DoFrGo:01}
Doucet, A., De~Freitas, N., and Gordon, N. (2001).
\newblock \emph{Sequential Monte Carlo Methods in Practice}, chapter An Introduction to Sequential Monte Carlo Methods.
\newblock Springer.

\bibitem[{Duník et~al.(2023)Duník, Matoušek, and Straka}]{DuMaSt:23}
{Duník, J., Matoušek, J., and Straka, O. (2023).
\newblock Design of efficient point-mass filter for linear and nonlinear dynamic models.
\newblock \emph{IEEE Control System Letters}, 42, 2005--2010.}

\bibitem[{Giraldo-Grueso et~al.(2024)Giraldo-Grueso, Popov, and Zanetti}]{GiPoZa:24}
Giraldo-Grueso, F., Popov, A.A., and Zanetti, R. (2024).
\newblock Gaussian mixture-based point mass filtering.
\newblock In \emph{2024 27th International Conference on Information Fusion (FUSION)}.

\bibitem[{H{\'e}non(1976)}]{He:76}
H{\'e}non, M. (1976).
\newblock A two-dimensional mapping with a strange attractor.
\newblock \emph{Communications in Mathemat. Physics}, 50, 69--77.

\bibitem[{Ma et~al.(2023)Ma, Ding, Li, and Fan}]{MaDiLiFa:23}
Ma, T., Ding, S., Li, Y., and Fan, J. (2023).
\newblock A review of terrain aided navigation for underwater vehicles.
\newblock \emph{Ocean Engineering}, 281, 114779.

\bibitem[{Matoušek et~al.(2025)Matoušek, Duník, and Straka}]{MaDuSt:25}
Matoušek, J., Duník, J., and Straka, O. (2025).
\newblock Lagrangian grid-based filters with application to terrain-aided navigation.
\newblock \emph{IEEE Signal Processing Magazine}, 42(2), 98--104.

\bibitem[{Orguner et~al.(2010)Orguner, Skoglar, Törnqvist, and Gustafsson}]{OrSkToGu:10}
Orguner, U., Skoglar, P., Törnqvist, D., and Gustafsson, F. (2010).
\newblock Combined point-mass and particle filter for target tracking.
\newblock In \emph{2010 IEEE Aerospace Conference}.

\bibitem[{Pace and Zhang(2011)}]{PaZh:11}
Pace, M. and Zhang, H. (2011).
\newblock Grid based phd filtering by fast Fourier transform.
\newblock In \emph{14th Internat. Conf. on Inf. Fusion}.

\bibitem[{Stoer and Bulirsch(1992)Stoer and Bulirsch}]{StBu:92}
{Stoer, J. and Bulirsch, R. (1992).
\newblock \emph{Introduction to Numerical Analysis}.
\newblock Springer-Verlag.}

\bibitem[{\v{S}imandl et~al.(2006)\v{S}imandl, Kr\'{a}lovec, and S\"{o}derstr\"{o}m}]{SiKraSo:06}
\v{S}imandl, M., Kr\'{a}lovec, J., and S\"{o}derstr\"{o}m, T. (2006).
\newblock Advanced point-mass method for nonlinear state estimation.
\newblock \emph{Automatica}, 42(7), 1133--1145.

\end{thebibliography}
                                                     % with bibtex (preferred)

\end{document}